\documentclass[aps,prd,twocolumn,showpacs,showkeys,nofootinbib,superscriptaddress]{revtex4-1}

\usepackage{longtable}
\usepackage{graphicx}
\usepackage{amsmath}
\usepackage{amsfonts}
\usepackage{amssymb}
\usepackage{natbib} 
\usepackage{setspace}
\usepackage{xspace}
\usepackage{cancel}
\usepackage{physics}
\usepackage{tensor,braket,color,slashed}
\usepackage{bbm}

\usepackage{txfonts,mathrsfs}
\usepackage{newtxtext}
\usepackage{newtxmath}

\graphicspath{{.}{figures/}}

\begin{document}

\title{
  \begin{flushright}
    \rightline{APCTP Pre2019-010} 
  \end{flushright}
  Kaon form factor in holographic QCD}

\author{Zainul Abidin}
\email{zainul.abidin@stkipsurya.ac.id}
\affiliation{Sekolah Tinggi Keguruan dan Ilmu Pendidikan Surya, Tangerang, Jawa Barat 15115, Indonesia}

\author{Parada~T.~P.~Hutauruk}
\email{parada.hutauruk@apctp.org}
\affiliation{Asia Pacific Center for Theoretical Physics, Pohang, Gyeongbuk 37673, Korea }

\date{\today}

\begin{abstract}
  The kaon form factor in the spacelike region is calculated using
  a holographic QCD model with the ``bottom-up'' approach. 
  We found that our result for the kaon form factor in low $Q^2$ has a remarkable agreement
  with the existing data, where $Q^2$ is the four-momentum transfer squared.
  The charge radius of the kaon as well as the kaon decay constant are
  found to be in a good agreement with the experiment data.
  We then predict the kaon form factor
  in the asymptotic region (larger $Q^2$) showing $1/Q^2$ behavior,
  which is consistent with the perturbative QCD prediction.  
\end{abstract}


\maketitle

\section{Introduction} 
\label{intro}

A theory of quantum chromodynamics (QCD), which is a non-Abelian gauge theory, is believed so far
as a correct theory of hadrons, where hadrons are the composite particles
made of quarks and gluons~\cite{CDG77}.
QCD has the essential features, namely, confinement and
chiral symmetry breaking~\cite{CDG77,MP77}. However, the form factor,
which is one of the nonperturbative quantities, is very difficult to compute directly from QCD.
Several theoretical and phenomenological models~\cite{HCT16,BWI94,Tandy97,SMBF12,KTT16}
as well as a lattice QCD calculation~\cite{KSDLL17} have been used
to calculate this nonperturbative quantity of QCD.

Apart from those models, during the past few years,
holographic QCD models, which are the complementary model of QCD, have also been applied to describe
the structure of hadrons, namely,  meson~\cite{KL07,GR07,GR07m,BEEGK03,GLSV11,AC08} and nucleon~\cite{GLSV11} form factors
as well as charmed meson~\cite{BKM17}, in order to gain a deep understanding of the structure of hadrons,
from a different substantially point of view. Surprisingly, these holographic models work well in predicting other hadron
observables, namely the decay constant and mass spectrum.
Also, one can argue that QCD approximately behaves as a conformal over a particular kinematic region~\cite{AC08,TB05}. 
Those holographic QCD models are able to preserve
confinement~\cite{GR07m,SS03,Polyakov97} and chiral symmetry breaking~\cite{GKK09,SS04,BEEGK03,GY04},
which is in many ways similar to the main properties of QCD in low energy,
after a few years since the holographic model was proposed~\cite{Maldacena97,Witten98}.

The original AdS/CFT correspondence~\cite{Maldacena97} has been first used to connect
a strongly coupled 4D conformal theory for large $N_c$, where $N_c$ is the color number,
and a weakly coupled gravity theory on AdS space. It then has been reconstructed starting from QCD
and its 5D gravity dual theory to reproduce the properties of QCD~\cite{KSS05,RP05,TB05,BT06}.

However, among those holographic models with various
approaches~\cite{KL07,GR07,GR07m,BEEGK03,GLSV11,AC08,BKM17,GR07m,SS03,Polyakov97,
  GKK09,SS04,BEEGK03,GY04,Maldacena97,Witten98},
only a few models have been used
to calculate the kaon form factor in a holographic QCD model with different approaches~\cite{BKM17}.
$K^{+} (u\bar{s})$ is a very interesting object, because
it consists of a strange quark, beside an up quark, where the mass of
the strange quark is heavier than the $u$ quark.
Experimentally, the existing data on the kaon form factor are very poor in higher $Q^2$,
only old data for low $Q^2$ are available~\cite{Amendolia86}, where $Q^2$ is the four-momentum transfer
squared. In the future, experiments will
measure the kaon form factor in higher $Q^2$~\cite{Carmignotto18,Horn017}. It would be
interesting to see how our complementary model, which is inspired by this AdS/CFT correspondence,
predicts the kaon form factor in higher $Q^2$. This work may pave the way to understand the strange quark properties
as well as the strange quark form factor.

In the present paper, we calculate the kaon form factor in holographic QCD, which is
a complementary approach of QCD. In this work, we adopt a ``bottom-up'' approach
of the AdS/CFT correspondence, instead of a ``top-down'' approach, where we employ the properties of QCD
to construct its 5D gravity dual theory as performed in Refs.~\cite{AC209,KSS05,RP05}.
We begin to describe the AdS/CFT correspondence formalism,
describing a correspondence between 4D operators $\cal{O}$(x) and fields in the 5D bulk $\phi$(x,z).
We then calculate the kaon form factor in holographic QCD.
We find the result on the kaon form factor
is in good agreement compared to the existing data in low $Q^2$~\cite{Amendolia86}.
We then predict the kaon form factor in higher $Q^2$.
Experimentally, the experimental data are really poor in higher $Q^2$.
We find that the kaon form factor in higher $Q^2$ is consistent with the perturbation QCD prediction~\cite{LB80}.
Next, we calculate the charge radius of the kaon in holography. We find that our result on the charge
radius is an excellent agreement with the data~\cite{Amendolia86} as well as the Particle Data Group (PDG)~\cite{Agashe14}.

This paper is organized as follows.
In Sec.~\ref{sec:formalism}, we briefly review the AdS/CFT correspondence, two- and three-point functions,
and how to extract the form factor of the kaon from holographic QCD in Sec.~\ref{sec:kaonEMFF}.
In Sec.~\ref{sec:chargeradius}, we present the calculation of the charge radius of the kaon.
In Sec.~\ref{sec:results}, numerical results are presented and
their implications are discussed.
Section~\ref{sec:summary} is devoted to a summary.
%
%
\section{Formalism} \label{sec:formalism}
%
\subsection{The AdS/QCD correspondence}
%

In this section, we briefly review the calculation of the vacuum expectation values
of the operators based on a generating function $Z_{4D}$ in the 4D space,
which is defined by
\begin{align}
  \label{eq:z4d}
  Z_{4D}[\phi^0] &= \left<\exp\left(i S_{4D}
  + i \int d^4x {\cal O}(x)\phi^0(x)\right)\right>\,,
\end{align}	
where $S_{4D}$ is the action for the $4D$ theory and $\phi^0$ is a source function together
with a specific operator ${\cal O}$ (x), which corresponds to the expectation value.
It then can be written by
\begin{align}
  \label{eq:vev}
  \left<0\left|{\cal T}{\cal O}(x_1)\ldots{\cal O}(x_n)\right|0\right> &=
  \frac{(-i)^n\delta Z_{4D}}{\delta \phi^0(x_1)\ldots\phi^0(x_n)}\,.
\end{align}

The following AdS/CFT correspondence provides the equivalence between the generating functional
of the connected correlation for the 4D theory and the effective partition function for the 5D theory:
\begin{align}
 \label{eq:adscft}
  Z_{4D} [\phi^0] &= \exp \left( iS_{5D} (\phi_{cl}) \right)\,,
\end{align}
where $\phi_{cl}$ is a solution of the 5D equation of motion with a boundary,
as defined in Eq.~(\ref{eq:boundary5D}).

We consider only the tree-level diagram on the 5D theory,
and we choose the following metric for the 5D space-time:
\begin{align}
  \label{eq:metric}
  ds^2 &= g_{MN} dx^M dx^N = \frac{1}{z^2}\left(\eta_{\mu\nu}dx^\mu dx^\nu - dz^2\right)\,,\quad \varepsilon<z<z_0\,,
\end{align}
where $x$ is the 4D space-time coordinate, $\eta_{\mu \nu} = \rm{diag} (1, -1, -1, -1)$ is the flat space metric,
and $z$ is the fifth coordinate,
which corresponds to the energy scale ($Q \sim 1/z$). We set $z = \varepsilon \rightarrow 0$
for the ultraviolet boundary of the 5D space that relates with the UV limit of QCD,
and the hard-wall cutoff at $z = z_0 = 1/\Lambda_{QCD}$ is the infrared boundary, which is used for
the conformal symmetry breaking of QCD.

The UV boundary value of the 5D field is the source of the corresponding 4D operator $\cal{O}$.
One can write the classical solution of the 5D field as
\begin{align}
  \label{eq:boundary5D}
  \phi_{cl}(x,z) &= \phi(x,z)\phi^0(x)\,.
\end{align}

The value of $\phi(x,\varepsilon) \rightarrow 1$ (or, in general, it goes to $\varepsilon^\Delta$).
Hence, $\phi^0(x)$ is identified as the UV-boundary value of the $\phi_{cl}(x,z)$ field.   
%

\subsection{The 5D AdS model}
%

The action in 5D theory is written as
\begin{align}
  \label{eq:action5D}
  S_{\rm{5D}} &= \int d^5x \sqrt{g}\, \textrm{Tr} \left\{\left|DX\right|^2+3\left|X\right|^2
  -\frac{1}{4g_5^2}\left(F_L^2+F_R^2\right)\right\},
\end{align}
where $g = |\textrm{det} g_{MN}|$ is the determinant of metric tensor,
$g_5$ is a gauge coupling parameter, which is fixed by the QCD operator product expansion,
and the bifundamental scalar field $X$
in Eq.~(\ref{eq:action5D}) is expressed by
\begin{align}
  \label{eq:biscalarfield}
  X (x,z)  &= \exp(i\pi^a(x,z) t^a)X^0(z)\exp(-i\pi^a(x,z)t^a),
\end{align}
where $t^a = \sigma^a /2$ are the SU($3$) generators with $\textrm{Tr}[t^a t^b] = \delta^{ab}/2$,
where $\sigma^a$ are the Pauli matrices.
The covariant derivative is defined as
\begin{align}
  \label{eq:codev}
  D_M X &= \partial_M X-iL_M X+iXR_M,
\end{align}
where the 5D space-time is denoted by the lowercase index of $M=(\mu,z)$ and
$F^L_{MN}$ is written as
\begin{align}
  \label{eq:Fmn}
  F^L_{MN} & =\partial_ML_N-\partial_N L_M-i[L_M,L_N],
\end{align}
Analogously for $F^R_{MN}$.

The $L$ and the $R$ fields can be written as vector field $V$ and the axial-vector field $A$:
\begin{align}
  L_M &= V_M+A_M\\
  R_M &= V_M-A_M.
\end{align}

In this work, we consider the following 4D operators that are defined by
the current operators $J^a_{L\mu} = \bar{\psi}_{qL} \gamma_\mu t^a \psi_{qL}$
and $J^a_{R\mu}=\bar{\psi}_{qR}\gamma_\mu t^a \psi_{qR}$ that correspond to the gauge fields
$L^a_\mu$ and $R^a_\mu$ in the 5D theory, respectively. The operator of $\bar{\psi}_{q_{R}}\psi_{q_{L}}$ corresponds
to a bifundamental scalar field $X^a$ in Eq.~(\ref{eq:biscalarfield}),
where the index $a = 1, 2, \ldots, 8$ for SU(3)
flavor symmetry, and index $\mu = 0,1,2,3$ for the space-time.
Note that the gauge invariance in the 5D theory is related with the global current conservation
in the 4D theory.

\subsection{Two-point functions}
%

Here we consider only the scalar parts of the action $X^0(z)$, up to second order, it gives
\begin{align}
  \label{eq:scalaraction}
  S_{\rm{scalar}} &= \int d^5x \sqrt{g}\,\textrm{Tr}\left(g^{MN}\partial_MX^0\partial_N X^0+3|X^0|^2\right),
\end{align}
where $g^{MN}$ is defined in Eq.~(\ref{eq:metric}), which is the nontrivial 5D metric.

The UV boundary of the scalar field $X^0$ is proportional to the quark mass matrix $M$,
which can be considered as the source for the operator of $\bar{\psi}_{R}\psi_{L}$.
Solving the equation of motion for the scalar field, it then gives
\begin{align}
  \label{eq:scalarsol}
  X^0(z) &= a_1 z+ a_3 z^3,
\end{align}
where $a_1$ is defined as in Ref.~\cite{AC209} by
\begin{align}
  a_1 &= \frac{1}{2}M \mathbbm{1} = \frac{1}{2}\left(\begin{array}{ccc}
    m_q&&\\
    &m_q&\\
    &&m_s
  \end{array}\right),
\end{align}
where we consider the SU(2) isospin symmetry, where the mass for the up and down quarks are identical.

Using the AdS/CFT correspondence, we then calculate
the quark condensate $\left<\bar{\psi}\psi\right> = \Sigma$ by
performing a functional derivative of the action in Eq.~(\ref{eq:scalaraction}),
evaluated on the classical solution, over $\delta M$ and identify 
\begin{align}
  \label{eq:condensate}
  a_3 &= \frac{1}{2}\Sigma \mathbbm{1} = \frac{1}{2}\left(\begin{array}{ccc}
    \sigma_q&&\\
    &\sigma_q&\\
    &&\sigma_s
  \end{array}\right).
\end{align}
We also assume that $\sigma_q=\sigma_s=\sigma$ and define $v_q(z)=m_q z +\sigma z^3$ and $v_s(z)=m_s z+\sigma z^3$.
%

\subsection{Transverse vector}
%

We now consider only vector parts of the action up to second order. It gives
\begin{align}
  \label{eq:vectoraction}
  S_{\rm{vector}} &= \int d^5x\, \sum_{a=4}^{8} \frac{1}{4g_5^2z} \left(-\left(\partial_MV^a_N-\partial_NV^a_M\right)^2
  +2\alpha^a(z)(V^a_M)^2\right).
\end{align}
A contraction over 5D metric $\eta_{ML}$ is implied. We then define
\begin{align}
  {\alpha^a}(z) =& \left \{ 
  \begin{array}{ll} 
    0 & a=1,2,3\\ 
    g_5^2\left(v_s-v_q\right)^2/(4z^2) \quad &  a=4,5,6,7\\ 
    0 &  a=8 \,,
  \end{array} \right .
\end{align}

We have gauge choice to set $V^a_z = 0$ except for $a=4,5,6,7$ because of
the nonzero (``mass term'') of the second term in the action of Eq.(\ref{eq:vectoraction}).
The equation of motion for the 4D Fourier transform of the transverse part of
the gauge field $V^a_{\perp,\mu}(q,z)$ is written as
\begin{align}
  \label{eq:vectoreom}
  \left(\partial_z\frac{1}{z}\partial_z +\frac{q^2-\alpha^a}{z}\right)V^a_{\perp,\mu}(q,z) &= 0,
\end{align}
where $\alpha^a=g_5^2(m_s-m_q)^2/4$ when $\sigma_s=\sigma_q$ for $a=4,5,6,7$.

One writes the transverse part of the vector field as $V^a_{\perp,\mu}(q,z)=V^{0,a}_{\perp,\mu}(q)V^a(q,z)$
with the so-called bulk-to-boundary propagator $V^a(q,z)$, which is
normalized to $V^a(q,\varepsilon)=1$ at the boundary condition $z=0$,
and $V^{0,a}_{\perp,\mu}(q)$ is the Fourier transform of the source of
the vector current $J^a_{V,\mu} = \bar{\psi}_{q_{v}} \gamma_\mu t^a \psi_{q_{v}}$
at the UV boundary $z=\varepsilon$.
We also impose a Neumann boundary condition $\partial_zV(q^2,z_0)=0$.
The solution for the bulk-to-boundary propagator is written as
\begin{align}
  \label{eq:transgaugefieldeom}
  V^a(q^2,z) &= \frac{\pi}{2} \tilde{q}z\left(\frac{Y_0(\tilde{q}z_0)}{J_0(\tilde{q}z_0)}J_1(\tilde{q}z)
  -Y_1(\tilde{q}z)\right),
\end{align}
where $\tilde{q}^2=q^2-\alpha^a$, and $Y_1 (x)$ and $J_1 (x)$ are the Bessel functions, respectively.

For spacelike four-momentum transferred $q^2=-Q^2<0$,
the solution in Eq.~(\ref{eq:transgaugefieldeom}) can be written as
\begin{align}
  \label{eq:solutionvectorspacelike}
  V^a(Q^2,z) &= \tilde{Q}z\left(\frac{K_0(\tilde{Q}z_0)}{I_0(\tilde{Q}z_0)}I_1(\tilde{Q}z)
  + K_1(\tilde{Q}z)\right),
\end{align}
where $\tilde{Q}=\sqrt{Q^2+\alpha^a}$, and $K_1 (x)$ and $I_1 (x)$ are the modified Bessel functions, respectively. 

The action on the solution in Eq.~(\ref{eq:solutionvectorspacelike}) is evaluated
with applying transverse projector
$\eta^{\mu\nu} \rightarrow \left(\eta^{\mu\nu}-\frac{q^\mu q^\nu}{q^2}\right) = P^{\mu \nu}_T$,
since $\partial_\mu V^{a,\mu}_\perp=0$, one has the form
\begin{align}
  \label{eq:actionvectortrans}
  S_{\rm{vector}} & =-\frac{1}{2g_5^2}\int \frac{d^4q}{(2\pi)^4} V^{0,a}_\mu(q)V^{0,a}_\nu(q)
  P^{\mu \nu}_T \frac{\partial_zV^a(q^2,\varepsilon)}{z}.
\end{align}

After solving the differential part, by the AdS/CFT correspondence,
we obtain the current-current two-point functions
\begin{align}
  \left<0\left|{\cal T}J^{a,\mu}_\perp(x) J^{b,\nu}_\perp(y)\right|0\right>
  &=\frac{i\delta^2 S_{5D}}{i^2\delta V^{0,a}_{\perp,\mu}(x)\delta V^{0,b}_{\perp,\nu}(y)},
\end{align}
where
\begin{align}
  V^{0,a}_\mu(q) &= \int d^4x e^{iqx}V^{0,a}_\mu(x),
\end{align}
and this leads to
\begin{align}
  i\int d^4x\, e^{iqx}\left<0\left|{\cal T}J^{a,\mu}_\perp(x) J^{b,\nu}_\perp(0)\right|0\right>  \nonumber \\
  = -\frac{1}{g_5^2}P^{\mu\nu}_T\delta^{ab}\frac{\partial_zV^a(q^2,\varepsilon)}{z},
\end{align}
where $\cal{T}$ is the time-ordering operator.

The bulk-to-boundary propagator can be written as
\begin{align}
  \label{eq:Vinpsi}
  V^a(q^2,z) &=\sum_{n=0}^{\infty} c^a_n(q^2) \psi_n(z),
\end{align}
where the wave function of $\psi_n$ satisfies the eigenvalue equation
\begin{align}
  \label{eq:normalizable}
  \left(\partial_z\frac{1}{z}\partial_z+\frac{\left(\left(M^a_{V,n}\right)^2
    -\alpha^a\right)}{z}\right)\psi^a_n(z) &= 0
\end{align}
which is normalized as
\begin{align}
  \int dz \frac{1}{z}\psi^a_n\psi^a_m &=\delta_{mn},
\end{align}
with boundary condition $\psi^a_n(\varepsilon)=0=\partial_z\psi^a_n(z_0)$ and the solution is
\begin{align}
  \psi^a_n(z) &= \frac{\sqrt{2}z\,J_1\left(z\sqrt{(M^a_{V,n})^2-\alpha^a}\right)}
      {z_0J_1\left(z_0\sqrt{(M^a_{V,n})^2-\alpha^a}\right)}
\end{align}
where the eigenvalues of $M^a_{V,n}$ (Kaluza-Klein tower of the mass of
the vector mesons: $\rho$ meson for $a=1,2,3$, $K^*$ meson for $a=4,5,6,7$,
and $\omega^0$ meson for $a=8$) are obtained from $J_0\left(z_0\sqrt{(M^a_{V,n})^2-\alpha^a}\right)=0$.

Using Eqs.~(\ref{eq:vectoreom}), (\ref{eq:Vinpsi}), and (\ref{eq:normalizable}),
we obtain
\begin{align}
  c^a_n(q^2) &= -\frac{\frac{1}{\varepsilon}\partial_z\psi_n(\varepsilon)}{q^2-(M^a_{V,n})^2}.
\end{align}

Since
\begin{align}
  i\int d^4x\, e^{iqx}\left<0\left|{\cal T}J^{a,\mu}_\perp(x) J^{b,\nu}_\perp(0)\right|0\right> \nonumber \\
  =\sum \frac{(f^a_{V,n})^2\delta^{ab}}{q^2-(M^a_{V,n})^2}\left(\eta^{\mu\nu}-\frac{q^\mu q^\nu}{q^2}\right)\nonumber\\
  +{\rm (nonpole\,terms)},
\end{align}
where the definition of $f^a_{V,n}$ is given by the matrix element of current,
$\left<0|J^a_{\mu}|V^c_n(q,\lambda)\right> = f^a_{V,n} \delta_{ac}\varepsilon_\mu(q,\lambda)$.
We identify $f^a_{V,n} =\partial_z\psi^a_n(\varepsilon)/\varepsilon$.
Also, the parameter $g_5^2 = 12\pi^2/N_c = 4\pi^2$ is fixed
from the quark bubble diagram in the leading order, with $N_c =3$ the number of color.
%

\subsection{Axial-vector and pseudoscalar}
%

The action for the axial-vector and pseudoscalar sector parts up to second order is
written as
\begin{align}
  S_{\rm{axial}} &= \int d^5x \sum_{a=1}^{8} \nonumber \\
  &\times \frac{1}{4g_5^2z}\left(-\left(\partial_MA^a_N-\partial_NA_M\right)^2
  + 2\beta^a(z)\left(\partial_M\pi^a-A_M^a\right)^2\right)
\end{align}
A contraction over 5D metric $\eta_{ML}$ is implied.
We have gauge choice $A^a_M \rightarrow A^a_M-\partial_M\lambda^a$,
and $\pi^a \rightarrow \pi^a -\lambda^a$ and $A^a_z=0$ are imposed.
We define
\begin{align}
  {\beta^a}(z) &= \left \{ 
  \begin{array}{ll} 
    g_5^2v_q^2/z^2 & a=1,2,3\\ 
    g_5^2(v_q+v_s)^2/(4z^2) \quad &  a=4,5,6,7\\ 
    g_5^2(v_q^2+2v_s^2)/(2z^2) &  a=8 \,,
  \end{array} \right .
\end{align}

For the field $\phi$ that comes from the longitudinal part, we define $A_{\parallel,\mu} = \partial_\mu \phi^a$.
We then write the Fourier transform of the fields in terms of the bulk-to-boundary propagators
that gives
\begin{align}
  \phi^a(p,z) &= \phi^a(p^2,z) \phi^{0a}(p) 
  = \phi^a(p^2,z) \frac{i p^\alpha}{p^2}
  A_{\parallel\alpha}^{0a}(p)	\,,
  \nonumber \\
  \pi^a(p,z) &= \pi^a(p^2,z) \frac{i p^\alpha}{p^2}
  A_{\parallel\alpha}^{0a}(p)\,,
  \nonumber \\
  A_{\perp \mu}^b(q,z) &= {\cal A}^b(q^2,z)\ 
  A_{\perp \mu}^{0b}(q)	
\end{align}
where $A_{\parallel\alpha}^{0a}(p)$ is the Fourier transform of the source function of
the 4D axial current operator $J^{a,\alpha}_{A,\parallel}$
and $A_{\perp \mu}^{0b}(q)$ is the Fourier transform of
the source function of the 4D axial current operator $J^{a,\alpha}_{A,\perp}$.

We obtain the coupled differential equations for
the longitudinal part of the axial-vector and pseudoscalar fields
as follows
\begin{align}
  -q^2\partial_z \phi^a(q^2,z)+\beta^a(z)\partial_z \pi^a(q^2,z)&=0\,,
  \label{phipi_profile1}	\\[1.25ex]
  \partial_z \left(\frac{1}{z}\partial_z\phi^a(q^2,z) \right)
  -\frac{\beta^a(z)}{z}\left(\phi^a(q^2,z)-\pi^a(q^2,z)\right)&=0\,,
  \label{phipi_profile2}
\end{align}
with the boundary conditions $\phi^a(q^2,\varepsilon)=0, \pi^a(q^2,\varepsilon)=-1$,
and $\partial_z\phi^a(q^2,z_0)=0=\partial_z\pi^a(q^2,z_0)$.
The expression for the transverse part of the axial-vector field is
analogous to the vector field. It then gives
\begin{align}
  \left(\partial_z\frac{1}{z}\partial_z+\frac{q^2-\beta^a(z)}{z}\right)A^a_{\perp}(q^2,z)=0.
  \label{equationaxialperp}
\end{align}

We then substitute the coupled differential in Eqs.~(\ref{phipi_profile1}) and~(\ref{phipi_profile2})
into a second-order equation, and we obtain
\begin{align}
  \partial_z\left(\frac{z}{\beta^a(z)}\partial_z y^a(q^2,z)\right)
  + z\left(\frac{q^2}{\beta^a(z)}-1\right)y^a(q^2,z) &=0\,.
  \label{axialcombi}
\end{align}
where $y^a(q^2,z)=\partial_z \phi^a(q^2,z)/z$.
In this form the boundary condition is $y(q^2,z_0)=0$
and $\varepsilon\partial_z y^a(q^2,\varepsilon)/\beta^a(\varepsilon)=1$.
We then have the solution as
\begin{align}
  y^a(q^2,z)=\sum \frac{(M^a_{\pi,n})^2y^a_n(\varepsilon) y^a_n(z)}{q^2-(M^a_{\pi,n})^2}
\end{align}
where $y^a_n(z)$ is a normalized solution of the eigenvalue in Eq.~(\ref{axialcombi})
with $q^2 = M^a_{\pi,n}$, boundary conditions $y_n(z_0)=0$,
and $\varepsilon\partial_z y^a_n(\varepsilon)/\beta(\varepsilon)=0$.
The normalization is
\begin{align}
  \int\frac{z}{\beta^a(z)}y^a_n(z)y^a_m(z)
  &=\frac{\delta_{mn}}{(M^a_{\pi,n})^2}
\end{align}

The eigenvalues of $(M^a_{\pi,m})^2$ are the Kaluza-Klein (KK) masses
for the pseudoscalar mesons: the pions for $a=1,2,3$, kaons for $a=4,5,6,7$,
and $\eta^0$'s for $a=8$.
The eigenvalues are obtained from the transverse part of the axial vector
in Eq.~(\ref{equationaxialperp}), giving us the KK mass of the $a_1$ and $K_1$ mesons.

As noted above, for the vector sector, we do not have the freedom to set $V^a_z=0$, for $a=4,5,6,7$.
However, if we define $V^a_z=-\partial_z\tilde{\pi}^a$,
$V^a_{\parallel,\mu}=\partial_\mu (\tilde{\phi}^a-\tilde{\pi}^a)$,
we obtain analogous equations as in Eqs.~(\ref{phipi_profile1}) and~(\ref{phipi_profile2})
with $\alpha^a(z)$ in place of $\beta^a(z)$.
We may proceed as above to obtain the eigenvalues of the KK mass of the scalar mesons $K^*_0$.

A current-current correlator for the axial sector is written as
\begin{align}
  \label{eq:correlatorAV}
  i\int_x e^{iqx} \big< 0 \big| \mathcal{T} 
  J^{a\mu }_{A\perp}(x)J^{b\nu }_{A\perp}(0) \big| 0 \big>
  &=-P^{\mu\nu}_T \delta^{ab}
  \frac{\partial_z {A}^a_\perp(q^2,\varepsilon)}{g_5^2 \varepsilon}\,,							\nonumber\\
  i\int_x e^{iqx}\big< 0 \big| \mathcal{T} 
  J^{a\mu}_{A\parallel}(x)J^{b\nu}_{A\parallel}(0) \big| 0 \big>
  &=-P^{\mu\nu}_L \delta^{ab}
  \frac{\partial_z \phi^a(q^2,\varepsilon)}{g_5^2 \varepsilon}\,,
\end{align}
where $P^{\mu\nu}_L=q^\mu q^\nu/q^2$. Using the completeness relation
$\sum_n \int \frac{d^3q}{(2\pi)^3 2 q^0} |n(q) \rangle \langle n(q)|= 1$ into
the correlators in Eq.~(\ref{eq:correlatorAV}), then multiplying $q^2-m_n^2$,
and taking a limit $q^2 \rightarrow m_n^2$,
one identifies the decay constant of the pseudoscalar mesons from AdS/QCD correspondence:
\begin{align}
  \label{eq:decayconstantAV}
  f^a_{A,n} &= -\frac{y^a_n(\varepsilon)}{g_5}
  = -\frac{\partial_z\phi^a_n(\varepsilon)}{g_5\varepsilon},
\end{align}
where the decay constants are defined by
\begin{align}
  \left<0\left|J^a_{A\mu\parallel}(0)\right|\pi^b_n(q)\right>=i f^a_{A,n} q_\mu \delta^{ab},
\end{align}
where the states of $|\pi^b_n(q)\big>$ are also considered
for the pions ($b=1,2,3$) as well as the kaons ($b=4,5,6,7$).
%

\section{Kaon electromagnetic form factor} \label{sec:kaonEMFF}
%

The electromagnetic form factors of the pion and kaon are presented in this section.
The relevant parts of the action are
\begin{align}
  S_{A_\parallel V_\perp A_\parallel}
  &=\int d^5x \bigg(\frac{1}{g_5^2z} \partial^\mu \phi^a\partial_\mu V^b_\nu\partial^\nu \phi^c f^{abc} \nonumber \\
  &+\frac{1}{z^3}\left(\partial^\mu\pi^a-\partial^\mu\phi^a\right)V^b_\mu \pi^c g^{abc} \nonumber\\
  &+\frac{1}{z^3}\left(-\frac{1}{2}\partial^\mu\left(\pi^a\pi^c\right)+\partial^\mu\phi^a \pi^c \right)V^b_\mu h^{abc}\bigg),
  \label{eq:actionAVA}
\end{align}
where the first term in Eq.~(\ref{eq:actionAVA}) that contains $f^{abc}$
arises from the gauge part of the original action
and other terms come from the chiral part. We then define
\begin{align}
  \label{eq:gabc}
  g^{abc} &= -2i \textrm{Tr} \left\{t^a,X_0 \right\}
  \left[ t^b, \left\{ t^c, X_0 \right\} \right],
  \nonumber \\
  h^{abc} &= -2i \textrm{Tr} \left[t^b,X_0 \right]
  \left\{ t^a, \left\{ t^c, X_0 \right\} \right\}.
\end{align}

If $g^{abc}$ and $h^{abc}$ in Eq.~(\ref{eq:gabc}) do not have $a$, $b$, or $c$,
which do not equal ``$8$'', it then gives
\begin{align}
  g^{abc} &= f^{abc} v_a v_c	\,,
  \nonumber \\
  h^{abc} &= f^{abc} (v_c - v_a) v_c	\,,
\end{align}
where for $X_0 = \frac{1}{2} c_0 + c_8 t^8$, the $v_a$ is defined as 
\begin{align}
  v_a = c_0 + c_8 d^{aa8} =
  \left\{	\begin{array}{cl}
    v_q \,, & a=1,2,3 \\
    \frac{1}{2}\left( v_q + v_s \right) \,, & a=4,5,6,7 \,.
  \end{array}
  \right.
\end{align}
where $f^{abc}$ and $d^{abc}$ are the structure constants of the SU($3$) algebra.

For three-point functions, we calculate three current operators
by taking the functional derivative of Eq.(\ref{eq:actionAVA}). One has
the form
\begin{align}
  \label{eq:3pf}
  &\langle 0 |{\cal T} J_{A\parallel}^{a,\alpha} (x) J_{\perp}^{\mu}(y) 
  J_{A\parallel}^{c,\beta}(w)	| 0 \rangle
  =\frac{i \delta S_{A_\parallel V_\perp A_\parallel} \qquad } 
  {i^3\delta A_{\parallel\alpha}^{0a}(x) \,
    \delta V_{\perp\mu}^{0b}(y) \, \delta A_{\parallel\beta}^{0c}(w)}
\end{align}

From Eq.~(\ref{eq:3pf}), we then extract the form factor using
the following matrix element
\begin{align}
  -f^{a*}_n f^b_m p^\beta k^\alpha\left<\pi^a_n(p)\left|J^{b,\mu}_\perp\right|\pi^c_m(k)\right>(2\pi)^4\delta^4(p-q-k) \nonumber\\
  = \lim_{ \stackrel{\scriptstyle{p^2\to (M_{\pi n}^a)^2}}{k^2\to (M_{\pi m}^c)^2}}
  \left(p^2-(M^a_{\pi n})^2\right)\left(k^2-(M^c_{\pi m})^2\right)\nonumber\\
  \times\int d^4x d^4y d^4w e^{i(px-qy-kw)} \nonumber \\
  \times \langle 0 |T J_{A\parallel}^{a\alpha} (x) J_{\perp}^{b\mu}(y)J_{A\parallel}^{c\beta}(w)| 0 \rangle.
\end{align}

We then obtain
\begin{align}
  \left<\pi^a_n(p)\left|J^{b,\mu}_\perp\right|\pi^c_m(k)\right>=&i(p+k)^\mu \int dz V^b(q^2,z)\frac{1}{z}\Big((\partial_z\phi^a_n)(\partial_z\phi^c_m)\nonumber\\
  &+\frac{g_5^2 v_av_c}{z^2}\left(\pi^a_n-\phi^a_n\right)\left(\pi^c_m-\phi^c_m\right)\Big)f^{abc}
\end{align}

For three quark flavors, the electromagnetic current operator is defined as
\begin{align}
  J_{EM,\mu} &= J^3_\mu+\frac{1}{\sqrt{3}}J^8_\mu
\end{align}

The current matrix element for the kaons $|K^+_n\rangle=|\pi^4_n+i\pi^5_n\rangle$ is
written as
\begin{align}
  \langle K^+_n(p_B)\left|J_{EM,\mu} \right|K^+_n(p_A)\rangle &= (p_A+p_B)^\mu F^K_{nn}(Q^2)
\end{align}
where $Q^2=-q^2=-(p_A-p_B)^2$ and a final expression for
the kaon form factor is obtained by
\begin{align}
  \label{eq:kaonff}
  F^K_{nn}(Q^2) &= \int dz V^3(Q^2,z)\frac{1}{z}\Big((\partial_z\phi^4_n)(\partial_z\phi^5_m) \nonumber \\
  &+\frac{g_5^2 v_4^2}{z^2}\left(\pi^4_n-\phi^4_n\right)\left(\pi^5_m-\phi^5_m\right)\Big).
\end{align}
%

\section{Kaon charge radius} \label{sec:chargeradius}
%

In this section, we present the charge radius of the kaon in low $Q^2$ as well as
in higher $Q^2$. For doing so,
we recall the kaon form factor in Eq.~(\ref{eq:kaonff}) that is
\begin{align}
  F^K_{nn}(Q^2) &= \int_0^{z_0} zV^a(Q^2,z) \rho^b_{nn}(z),
\end{align}
where $a=1,2,3$, $b=4,5,6,7$, and
$\rho^b_{nn} (z)$ is defined by
\begin{align}
  \rho^b_{nn}(z) &=\frac{(\partial_z\phi^b_n)^2}{z^2}+\frac{g_5^2 v_b^2}{z^4}\left(\pi^b_n-\phi^b_n
  \right)^2.
  \label{rhoz}
\end{align}

In the limit of $Q\rightarrow 0$, the bulk-to-boundary propagator
in Eq.(\ref{eq:solutionvectorspacelike}) is written as,
\begin{align}
  V^a(Q^2,z)=1-\frac{Q^2z^2}{4}\left(1-2\ln\left(\frac{z}{z_0}\right)\right).
\end{align}

Using the expansion of Eq.(\ref{eq:solutionvectorspacelike}),
we obtain the radius of the kaon as follows
\begin{align}
  \langle {r^2_{Kn}}\rangle =-6\frac{dF^K_{nn}(Q^2)}{d(Q^2)}
  &=\int_0^{z_0} \frac{6}{4}z^3 \left(1-2\ln\left(\frac{z}{z_0}\right)\right)\rho^b_{nn}(z).
  \label{radiusK}
\end{align}
%

\section{Numerical Results} \label{sec:results}
%

Our numerical results for the kaon masses, decay constants, and kaon form factors
are presented in this section. Following Ref.~\cite{AC08}, we fix the parameter
values of the hard-wall cutoff at $z_0=(322.5~{\rm MeV})^{-1}$, which is chosen
to fit the lightest $\rho$ meson mass $M^a_{V,1}=775.5$ MeV for $a=1, 2, 3$.
Parameters $m_q$ and $\sigma_q$ is chosen to reproduce the pion mass and decay constant,
respectively.
Given the values of the pion mass $M^a_{\pi,1} = 139.6$ MeV,
and decay constant $f^a_1 = 92.4$ MeV for $a=1,2,3$, respectively,
we obtain the light current quark mass $m_q = 2.29 $ MeV
and condensate $\sigma_q = (328.3\,{\rm MeV})^3$.
We then fix $\sigma_s = \sigma_q$. The strange current quark mass $m_s=51.96$ MeV is
chosen to fit the kaon mass $M^a_{K,1} = 495.7$ MeV for $a=4, 5, 6, 7$
(the masses for the $K^+$, $K^-$, $K^0$, and $\bar{K}^0$, respectively).
We simply consider $m_q$, $m_s$, and $\sigma$ as model parameters,
not the (realistic) physical values of the quark mass and quark condensate.
For getting a better connection between the light current quark mass and condensate,
we redefine the parameters by taking $m_q\rightarrow \sqrt{N_c}/2\pi$,
and $\sigma\rightarrow 2\pi/\sqrt{N_c}$ without modifying
the above results for the two-point and three-point functions.
With this redefinition, we obtain $m_q= 8.31$ MeV, $m_s=188.5$ MeV,
and $\sigma=(213.7\,{\rm MeV})^3$.

Using the obtained parameters above, we determine the decay constant of
the lightest KK of the kaons $f_{K^{+}} = 104$ MeV, and the mass
and decay constant of the $K_0^*$ are
$m_{K_0^*}=791$ MeV and $f_{K_0^*}=28$ MeV, respectively.
The decay constant of the $\rho$ meson $f_\rho^{1/2}=329$ MeV.
The mass and decay constant of the lightest KK of the vector mesons $K^*$ are
$m_{K^*} = 791$ MeV, and $f_{K^*}^{1/2} = 329$ MeV, respectively.
For the axial vector mesons, the mass and decay constant of the $a_1$
are $m_{a_1}=1366$ MeV, and $f_{a_1}^{1/2}=489$ MeV, respectively.
For the $K_1$, we obtain $m_{K_1}=1458$ MeV, and $f_{K_1}^{1/2}=511$ MeV. 
The values of the decay constant and the mass of the kaon obtained are consistent with
PDG~\cite{Agashe14}.

\begin{figure}[t]
  \centering\includegraphics[width=0.95\columnwidth]{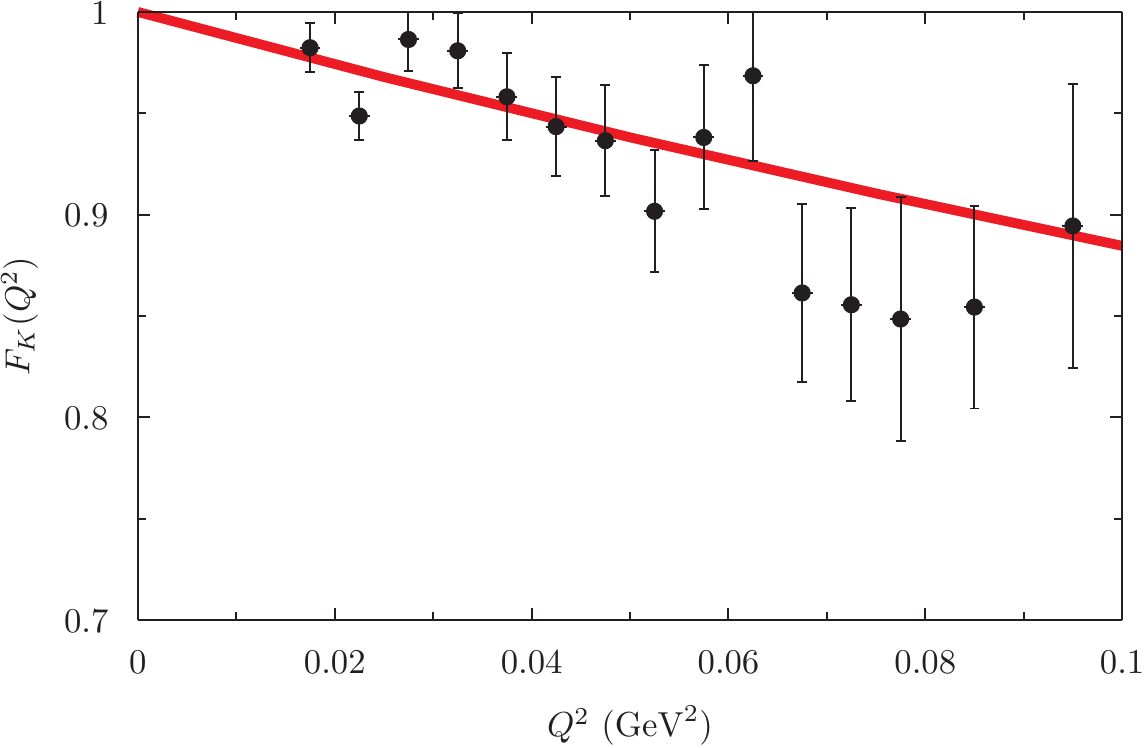}
  \caption{\label{fig1} The kaon form factor (solid line) compared to the
    existing data taken from Ref.~\cite{Amendolia86}.}
\end{figure}

\begin{figure}[t]
  \centering\includegraphics[width=0.95\columnwidth]{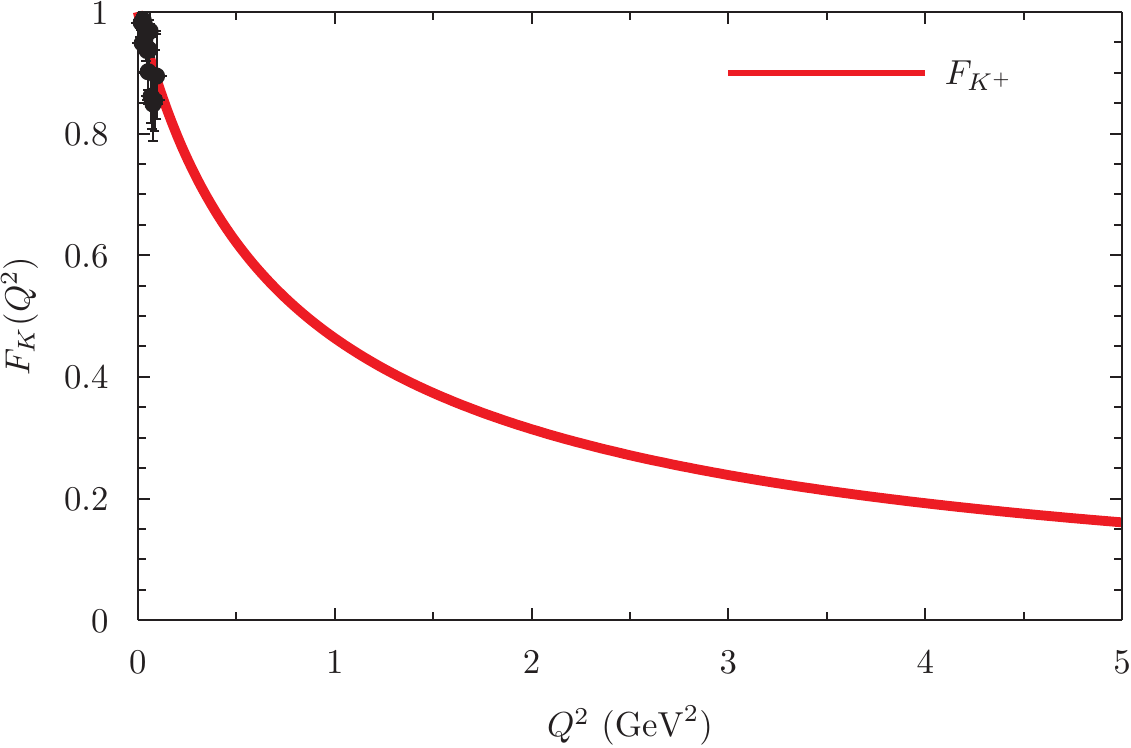}
  \caption{\label{fig2}. The kaon form factor (solid line) compared with data at low $Q^2$.
    The experiment data are taken from Ref.~\cite{Amendolia86}. }
\end{figure}

\begin{figure}[t]
  \centering\includegraphics[width=0.95\columnwidth]{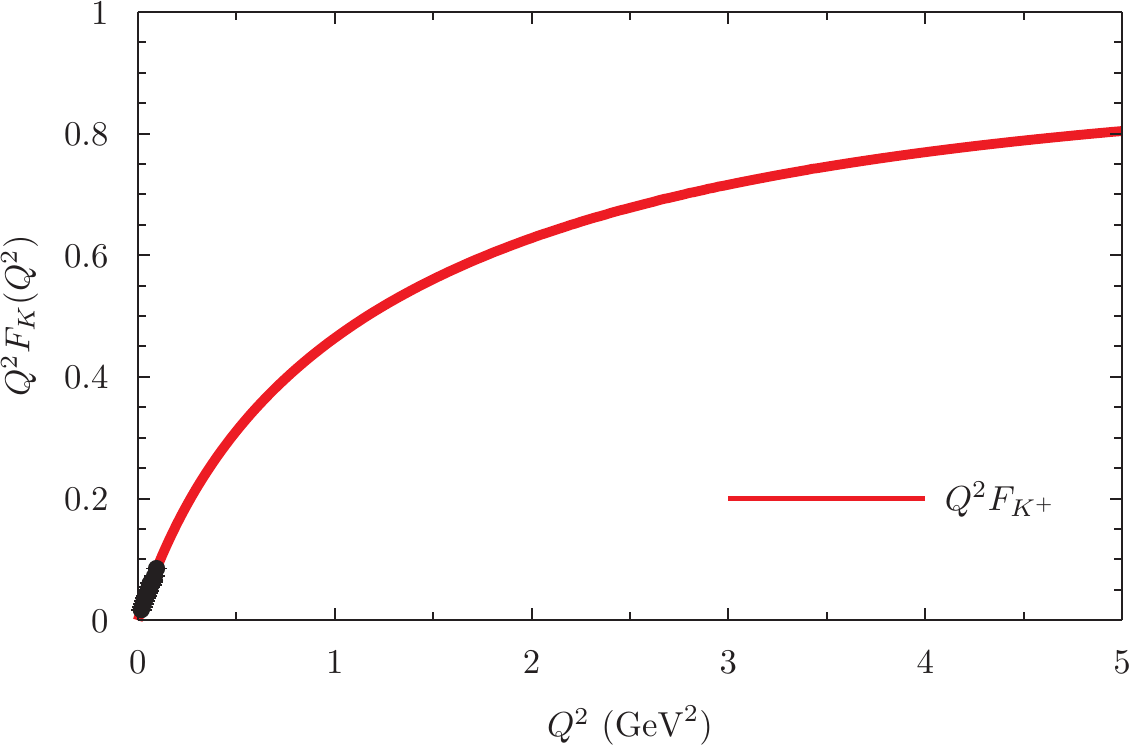}
  \caption{\label{fig3}. The same as in Fig.~\ref{fig1} but for $Q^2 F_K^{+} (Q^2)$.
    The experiment data are taken from Ref.~\cite{Amendolia86}.}
\end{figure}

Results for the kaon form factor are shown in Figs.~\ref{fig1}-~\ref{fig3}.
Figure~\ref{fig1} shows our prediction for the kaon form factor compared to
the existing data~\cite{Amendolia86} in low $Q^2$. We find that our prediction
is in excellent agreement with the data~\cite{Amendolia86}. 
We then calculate the kaon form factor up to $Q^2 =$ 5 $\textrm{GeV}$ to
anticipate the higher $Q^2$ data which will collect soon~\cite{Carmignotto18,Horn017},
as in Fig.~\ref{fig2}, however, experimentally, the kaon form factor is poorly known. 

Figure~\ref{fig3} shows the same results as in Fig.~\ref{fig2}, but for $Q^2F_K(Q^2)$.
For larger $Q^2$ (asymptotic region), the bulk-to-boundary propagator is written as
\begin{align}
  V^a(Q^2,z)\stackrel{Q^2\to \infty}{=} (Qz) K_1(Qz)\approx \sqrt{\frac{\pi Qz}{2}}e^{-Qz},
\end{align}
which goes to zero unless $z$ is infinitesimal, $z\sim 1/Q$.
Note that the first term in Eq.~(\ref{rhoz}) goes to $g_5^2(f^a_n)^2$ when $z\to 0$,
while the second term goes like $\varepsilon^2\to 0$. The quantity $zV^a(Q^2,z)$ behaves
like a delta function picking up $\rho^a_{nn}(z)$ at $z\to 0$. The upper limit of the form factor integral
can be set to infinity as the integrated vanish at large $z$. Then, the kaon form factor
in higher $Q^2$ is defined by
\begin{align}
  F^a(Q^2)&\stackrel{Q^2\to \infty}{=}
  \frac{g_5^2 (f^a_{n})^2}{Q^2}\int_{0}^{\infty}dw\, w^2 K_1(w)\nonumber\\
  &\quad=\frac{2g_5^2 (f^a_{n})^2}{Q^2}=\frac{8\pi^2(f^a_{n})^2}{Q^2}.
\end{align}

We find that the kaon form factor for larger $Q^2$ agrees well
with the perturbative QCD prediction~\cite{LB80}.

Using Eq.~(\ref{radiusK}), we obtain the charge radius
for the lightest kaon $r_{K^{+}} =$ 0.56 $\,\textrm{fm}$.
We find that our result is in excellent agreement with the experimental data~\cite{Amendolia86} and PDG~\cite{Agashe14}.

We also compare our model approach with the work of Ref.~\cite{Ahmady:2018muv},
which uses a light-front (LF) holographic approach,
where the holographic expression in 5D AdS space is matched to QCD in the LF frame.
In this approach, to incorporate the quark mass,
the ``effective potential'' in the AdS space is modified by adding a term
to obtain the meson mass expression matches with the quark mass contribution in LF QCD.
Contrary to this approach, we introduce the quark mass parameter,
as a source of the quark bilinear operator $\bar{\psi}_{R}\psi_{L}$ in the AdS boundary,
which is consistent with the AdS/CFT rule, and it appears as a coefficient in the background field, $X_0$.
Consequently, the quark mass parameter appears differently in the effective potential,
compared to the work of Ref.~\cite{Ahmady:2018muv}.
In addition, they identify the light-front wave function, where hadron properties
are encoded, by comparing the electromagnetic form factor in AdS and the LF QCD form factor.

In comparing our obtained results with their results on the kaon form factors,
their results for the charge radius of the $K^+$ are slightly larger than our result in the low $Q^2$ regime,
where the charge radius is $0.615$ fm for the dynamical spin parameter $B=0$ and even larger for $B>0$.
However, the behavior prediction of the kaon form factor in the large $Q^2$ regime, which goes like $1/Q^2$,
is similar to our obtained result.

We note that, in this paper, we started with an AdS Lagrangian that has $SU(3)_L \times SU(3)_R$
symmetry and it reproduces a chiral symmetry breaking of QCD. An approximate relation due to a chiral-symmetry-breaking-like, Gell-Mann-Oakes-Renner relation is preserved in our approach.

%

%

\section{Summary} \label{sec:summary}
%

In summary, we have computed the kaon form factor in holographic QCD, which is
a complementary approach of QCD. We adopt a ``bottom-up'' approach
of the AdS/CFT correspondence, instead of a ``top-down'' approach,
where we employ the properties of QCD
to construct its 5D gravity dual theory.
We begin to describe the AdS/CFT correspondence formalism,
describing a correspondence between 4D operators $\cal{O}$(x)
and fields in the 5D bulk $\phi$(x,z).
We calculate the kaon form factor in holographic QCD.

The result for the kaon form factor is in good agreement with
the existing data in low $Q^2$. We then predict the kaon
form factor in higher $Q^2$. We found that the kaon form factor
in higher $Q^2$ is consistent with the perturbation QCD prediction.

We finally calculate the charge radius of the kaon in holography QCD.
We obtained $r_K^{+} =$ 0.56 $\rm{fm}$, which is in excellent agreement with the data
as well as the Particle Data Group.
In the future, it would be interesting to extend the calculation of
the form factor and gravitational form factor of the B and D mesons, which
contain the bottom and charm quarks, respectively, using the holographic QCD model.

%
\begin{acknowledgments} 
  The work of P.T.P.H. was supported by the Ministry of Science, Information Communication and Technology and Future Planning,
  Gyeongsangbuk-do and Pohang City through the Young Scientist Training Asia-Pacific Economic
  Cooperation program of Asia Pacific Center for Theoretical Physics (APCTP). 
\end{acknowledgments}
%



\begin{thebibliography}{10}

\bibitem{CDG77} 
C.~G.~Callan, Jr., R.~F.~Dashen and D.~J.~Gross,
\newblock Toward a theory of the strong interactions,
\newblock Phys.\ Rev.\ D {\bf 17}, 2717 (1978).

\bibitem{MP77} 
W.~J.~Marciano and H.~Pagels,
\newblock Quantum chromodynamics: A review,
\newblock Phys.\ Rep.\  {\bf 36}, 137 (1978).

\bibitem{HCT16} 
P.~T.~P.~Hutauruk, I.~C.~Cloet and A.~W.~Thomas,
\newblock Flavor dependence of the pion and kaon form factors and parton distribution functions,
\newblock Phys.\ Rev.\ C {\bf 94}, 035201 (2016).

\bibitem{BWI94} 
W.~W.~Buck, R.~A.~Williams and H.~Ito,
\newblock Elastic charge form-factors for K mesons,
\newblock  Phys.\ Lett.\ B {\bf 351}, 24 (1995).

\bibitem{Tandy97} 
P.~C.~Tandy,
\newblock Hadron physics from the global color model of QCD,
\newblock  Prog.\ Part.\ Nucl.\ Phys.\  {\bf 39}, 117 (1997).

\bibitem{SMBF12} 
E.~O.~da Silva, J.~P.~B.~C.~de Melo, B.~El-Bennich and V.~S.~Filho,
\newblock Pion and kaon elastic form factors in a refined light-front model,
\newblock Phys.\ Rev.\ C {\bf 86}, 038202 (2012).

\bibitem{KTT16} 
A.~F.~Krutov, S.~V.~Troitsky and V.~E.~Troitsky,
\newblock The $K$-meson form factor and charge radius: Linking low-energy data to future Jefferson Laboratory measurements,
\newblock  Eur.\ Phys.\ J.\ C {\bf 77}, 464 (2017).

\bibitem{KSDLL17} 
J.~Koponen, A.~Zimermmane-Santos, C.~Davies, G.~P.~Lepage and A.~Lytle,
\newblock Light meson form factors at high $Q^2$ from lattice QCD,
\newblock  EPJ Web Conf.\  {\bf 175}, 06015 (2018).

\bibitem{KL07} 
H.~J.~Kwee and R.~F.~Lebed,
\newblock Pion form factor in improved holographic QCD backgrounds,
\newblock  Phys.\ Rev.\ D {\bf 77}, 115007 (2008).

\bibitem{GR07} 
H.~R.~Grigoryan and A.~V.~Radyushkin,
\newblock Pion form-factor in chiral limit of hard-wall AdS/QCD model,
\newblock  Phys.\ Rev.\ D {\bf 76}, 115007 (2007).

\bibitem{GR07m} 
H.~R.~Grigoryan and A.~V.~Radyushkin,
\newblock Structure of vector mesons in holographic model with linear confinement,
\newblock Phys.\ Rev.\ D {\bf 76}, 095007 (2007).

\bibitem{BEEGK03} 
J.~Babington, J.~Erdmenger, N.~J.~Evans, Z.~Guralnik and I.~Kirsch,
\newblock Chiral symmetry breaking and pions in nonsupersymmetric gauge / gravity duals,
\newblock  Phys.\ Rev.\ D {\bf 69}, 066007 (2004).

\bibitem{AC08} 
Z.~Abidin and C.~E.~Carlson,
\newblock Gravitational form factors of vector mesons in an AdS/QCD model,
\newblock Phys.\ Rev.\ D {\bf 77}, 095007 (2008).

\bibitem{GLSV11} 
T.~Gutsche, V.~E.~Lyubovitskij, I.~Schmidt and A.~Vega,
\newblock Dilaton in a soft-wall holographic approach to mesons and baryons,
\newblock  Phys.\ Rev.\ D {\bf 85}, 076003 (2012).

\bibitem{BKM17} 
A.~Ballon-Bayona, G.~Krein and C.~Miller,
\newblock Strong couplings and form factors of charmed mesons in holographic QCD,
\newblock  Phys.\ Rev.\ D {\bf 96}, 014017 (2017).

\bibitem{TB05} 
G.~F.~de Teramond and S.~J.~Brodsky,
\newblock Hadronic Spectrum of a Holographic Dual of QCD,
\newblock Phys.\ Rev.\ Lett.\  {\bf 94}, 201601 (2005).

\bibitem{SS03} 
T.~Sakai and J.~Sonnenschein,
\newblock Probing flavored mesons of confining gauge theories by supergravity,
\newblock J. High Energy Phys. {\bf 0309} (2003) 047.

\bibitem{Polyakov97} 
A.~M.~Polyakov,
\newblock String theory and quark confinement,
\newblock Nucl.\ Phys.\ Proc.\ Suppl.\  {\bf 68}, 1 (1998).
  
\bibitem{GKK09} 
T.~Gherghetta, J.~I.~Kapusta and T.~M.~Kelley,
\newblock Chiral symmetry breaking in the soft-wall AdS/QCD model,
\newblock  Phys.\ Rev.\ D {\bf 79}, 076003 (2009).

\bibitem{SS04} 
T.~Sakai and S.~Sugimoto,
\newblock Low energy hadron physics in holographic QCD,
\newblock  Prog.\ Theor.\ Phys.\  {\bf 113}, 843 (2005).

\bibitem{GY04} 
K.~Ghoroku and M.~Yahiro,
\newblock Chiral symmetry breaking driven by dilaton,
\newblock Phys.\ Lett.\ B {\bf 604}, 235 (2004).
  
\bibitem{Maldacena97} 
J.~M.~Maldacena,
\newblock The large N limit of superconformal field theories and supergravity,
\newblock Int.\ J.\ Theor.\ Phys.\  {\bf 38}, 1113 (1999).
\newblock [Adv.\ Theor.\ Math.\ Phys.\  {\bf 2}, 231 (1998)].

\bibitem{Witten98} 
E.~Witten,
\newblock Anti-de Sitter space and holography,
\newblock Adv.\ Theor.\ Math.\ Phys.\  {\bf 2}, 253 (1998).

\bibitem{KSS05} 
J.~Erlich, E.~Katz, D.~T.~Son and M.~A.~Stephanov,
\newblock  QCD and a Holographic Model of Hadrons,
\newblock  Phys.\ Rev.\ Lett.\  {\bf 95}, 261602 (2005).

\bibitem{RP05} 
L.~Da Rold and A.~Pomarol,
\newblock Chiral symmetry breaking from five dimensional spaces,
\newblock  Nucl.\ Phys.\ {\bf B721}, 79 (2005).

\bibitem{BT06} 
S.~J.~Brodsky and G.~F.~de Teramond,
\newblock Hadronic Spectra and Light-Front Wavefunctions in Holographic QCD,
\newblock Phys.\ Rev.\ Lett.\  {\bf 96}, 201601 (2006).

\bibitem{Amendolia86} 
S.~R.~Amendolia {\it et al.},
\newblock A measurement of the kaon charge radius,
\newblock  Phys.\ Lett.\ B {\bf 178}, 435 (1986).

\bibitem{Carmignotto18} 
M.~Carmignotto {\it et al.},
\newblock Separated kaon electroproduction cross cection and the kaon form factor from 6 GeV JLab data,
\newblock Phys.\ Rev.\ C {\bf 97}, 025204 (2018).

\bibitem{Horn017} 
T.~Horn,
\newblock Meson Form Factors and Deep Exclusive Meson Production Experiments,
\newblock  EPJ Web Conf.\  {\bf 137}, 05005 (2017).

\bibitem{AC209} 
Z.~Abidin and C.~E.~Carlson,
\newblock Strange hadrons and kaon-to-pion transition form factors from holography,
\newblock Phys.\ Rev.\ D {\bf 80}, 115010 (2009).

\bibitem{LB80} 
G.~P.~Lepage and S.~J.~Brodsky,
\newblock Exclusive processes in perturbative quantum chromodynamics,
\newblock Phys.\ Rev.\ D {\bf 22}, 2157 (1980).

\bibitem{Agashe14} 
K.~A.~Olive {\it et al.} (Particle Data Group),
\newblock Review of Particle Physics,
\newblock Chin.\ Phys.\ C {\bf 38}, 090001 (2014).

\bibitem{Ahmady:2018muv} 
M.~Ahmady, C.~Mondal and R.~Sandapen,
\newblock Dynamical spin effects in the holographic light-front wavefunctions of light pseudoscalar mesons,
\newblock Phys.\ Rev.\ D {\bf 98}, 034010 (2018)
  
\end{thebibliography}
\end{document}